# Homogeneous Vapor Nucleation Using Isothermal Multibody Dissipative Particle Dynamics

Anuj Chaudhri[a] and Jennifer R. Lukes[b]

Department of Mechanical Engineering and Applied Mechanics, University of Pennsylvania,

Philadelphia, PA 19104 USA

Homogeneous nucleation of a vapor bubble from a bulk metastable phase is studied here for the first time using the multibody dissipative particle dynamics model. The van der Waals equation of state is used to define a conservative force that describes the model fluid used in this study. Additional computational tools were developed to help visualize the vapor bubble formed in the system. Continuum thermodynamic ideas are used to help identify the bubble interface and separate particles that constitute the vapor bubble and bulk liquid regions respectively. Using this idea, the density variation is found to vary smoothly from the liquid to vapor regions in the system. The square-gradient term that arises from the free energy expansion of the density is found to have no significant effect on the radial local density variation.

## I. INTRODUCTION

Phase change processes play a pivotal role in virtually all technological applications such as refrigeration, petroleum and chemical processing, electronics cooling, power plants, climatology and spacecraft environments [1]. The heat transfer and fluid flow processes associated with liquid-vapor phase change phenomena are among the most complex transport conditions that are encountered in engineering applications. The complexities typically arise from the interplay of multiple length and time scales involved in the process, non-equilibrium effects, and other effects arising from dynamic interfaces. Nucleation of one phase into another occurs at the atomic scales whereas growth of the nuclei and their interaction with the system occurs at much longer length and time scales [2]. Phase change processes associated with boiling and condensation are perhaps the most important since they are associated with a diverse set of applications such as power and refrigeration cycles, electronics cooling for aircraft and ship systems, and natural weather cycles [3-4]. For example, boiling is a complex nonequilibrium multiscale process, where the physics is linked from nano to meso to macro scales. Nucleate boiling is associated with vapor bubbles being formed at the heater surface, bubble merger and breakup that leads to enhancements in heat transfer. For each of these processes, the physics from different scales have to be seamlessly connected. The long term goal is to be able to solve the phase change heat transfer problems by solving basic equations similar to single phase flows [5]. Hence, there is a critical need for new modeling paradigms that are capable of capturing fundamental liquid-vapor phase change behavior at each scale (nano, meso and macro) and extracting practical information such as macroscopic observables from them. Advances in numerical techniques at the nano and meso scales have a lot to offer in terms of visualizing and understanding nucleation processes in liquid-vapor systems.

[a] Presently at University of Chicago; Electronic mail: chaudhri@uchicago.edu

[b] Electronic mail: jrlukes@seas.upenn.edu

All the studies using Molecular Dynamics (MD) have focused on single bubble nucleation [6-12] and have been extremely useful to understand liquid/vapor phase change at the atomic scale. These are capable of revealing fundamental physical properties like local temperature, surface tension, density, and fluid structure. Okumura and Ito [13] have modeled the dynamics of bubbles using nonequilibrium methods. However MD studies are limited to smaller length and time scales. Mesoscopic methods such as Lattice Boltzmann (LB) and Lattice Gas (LG) have had limited success with modeling these problems. Shan and Chen [14] were the first to introduce a multiphase model and use it to model liquid-gas interfaces and bubble nucleation. The method has been used to model planar interfaces and drops in two phase flows and recently for bubble formation in shearing flows. The major drawbacks of the LB method come from the lack of a complete multiphase thermal LB model [15], although there is some progress in that direction [16]. In addition, the LG and LB models do not include fluctuations that are important at micro/meso scales [17]. Dissipative particle dynamics (DPD) offers a good alternative to the grid-based LG and LB methods.

Dissipative particle dynamics (DPD) is a meshless, coarse-grained, particle-based mesoscopic method that has been used to model complex fluid systems such as lipid bilayer membranes, vesicles, micelles, binary immiscible fluids, suspensions, composites and polymersomes [18-19]. The equations of DPD are stochastic differential equations with conservative, dissipative and random forces. The key component of DPD that advances the DPD particle ('bead') velocities over time are the three forces:

$$\mathbf{F}_i = \mathbf{F}_i^{\mathrm{C}} + \mathbf{F}_i^{\mathrm{D}} + \mathbf{F}_i^{\mathrm{R}} \qquad (1)$$

The conservative $(\mathbf{F}^{\mathrm{C}})$, random $(\mathbf{F}^{\mathrm{R}})$, and dissipative $(\mathbf{F}^{\mathrm{D}})$ forces depend on position and the dissipative forces additionally depend on the velocities of the interacting pairs of beads. The conservative force is usually chosen as soft repulsive as proposed by Groot and Warren in [20]:

$$\mathbf{F}_i^C = \sum_{j \neq i} a_{ij} w_{ij}^C(\mathbf{r}_{ij}) \hat{\mathbf{r}}_{ij} \qquad (2)$$

$$\hat{\mathbf{r}}_{ij} = \frac{\mathbf{r}_i - \mathbf{r}_j}{|\mathbf{r}_i - \mathbf{r}_j|} = \frac{\mathbf{r}_i - \mathbf{r}_j}{|\mathbf{r}_{ij}|} \qquad (3)$$

Here $a_{ij}$ represents the strength of the conservative force, and $w_{ij}^C(\mathbf{r}_{ij})$ is a weight function for the conservative force. The exact expressions for the dimensionless equations and their scaling factors can be found in [21]. DPD includes fluctuations, conserves linear and angular momentum and has a thermal model also, which makes it an attractive method to use.

The isothermal DPD model is good at handling immiscible liquid-liquid phase separation [22] and liquid-solid phase separation [23] but is unable to handle the liquid-vapor interface due to the unusual equation of state. The conservative force expression most commonly used in DPD gives rise to a quadratic dependence of density on pressure:

$$\bar{p} = \bar{\rho} \overline{k_B T} + c \bar{a} \bar{\rho}^2 \qquad (4)$$

Here $\bar{p}$ is the pressure of the system, $\bar{\rho}$ is the dimensionless density, $\bar{a}$ is the strength of the conservative force and $c$ is a constant. Equation (4) shows that $\frac{d\bar{p}}{d\bar{\rho}} > 0$ for all $\bar{\rho}$. The monotonic increase of pressure with density arising from Eq. (4) precludes liquid-vapor phase transitions in traditional DPD.

To overcome the limitation of the conservative force and the pressure equation of state in Eq. (4), Pagonabarraga et al. [24] proposed a new definition of the conservative force inspired from density functional theory of inhomogeneous fluids [25-27]. Their new version is called multibody DPD (MDPD), where the conservative force is now dependent on the instantaneous local particle density of the fluid. They investigated a modified van der Waals model and a binary mixture system in two dimensions. This method has been modified to include particle correlations by Trofimov et al. [28]. Another model has been proposed by Warren [29] to incorporate phase change by employing conservative forces that have different cut-off radii for the attractive and repulsive components of the forces. This model was used to study planar interfaces and pendant drop simulations. Tiwari et al. [30] have developed a model based on the multibody method by adding an additional term to the conservative force that depends on the density gradients. It has been used to model two phase flow problems such as capillary waves and oscillations of a liquid cylinder. More details on the multibody formalism can be found in [24]. In this study, the van der Waals free energy model was picked to define the free energy density and the conservative force can be defined as [30]:

$$\mathbf{F}_i^{\mathrm{C}} = -\sum_{j=1}^{N} \left\{ \left( \frac{bk_BT}{1-bn_i} - a \right) + \left( \frac{bk_BT}{1-bn_j} - a \right) \right\} \frac{w'}{w} r_{ij} \hat{r}_{ij} \qquad (5)$$

Using the scaling factors defined in [21] and defining additional scaling factors for $a$ and $b$ as $a^*$ and $b^*$, the dimensionless form of the conservative force in Eq. (5) can be written as:

$$\overline{\mathbf{F}}_i^{\mathrm{C}} = -\sum_{j=1}^{N} \left\{ \left( \frac{\overline{b}\,\overline{k_BT}}{1-\overline{b}\,\overline{n}_i} - \overline{a} \right) + \left( \frac{\overline{b}\,\overline{k_BT}}{1-\overline{b}\,\overline{n}_j} - \overline{a} \right) \right\} \frac{\overline{w}'}{\overline{w}} \overline{r}_{ij} \hat{\overline{r}}_{ij} \qquad (6)$$

The scale factor $b^*$ can be obtained by equating the dimensionless grouping $b^*n^*$ to one. Since $n^*$ has units of inverse volume $1/[r^*]^3$, $b^*$ is identified as $b^* = [r^*]^3$, where $r^*$ is the scale factor for length. The scaling factor for force is given in the units of $k_BT^*/r^*$ from [21].

Using this the scale factor for $a^*$ can be identified as $a^* = k_B T^* [r^*]^3$. Similarly, the dimensionless counterparts of $n$ and $w(r_{ij})$ can be defined easily.

In the density functional theory expansion of the free energy, the contribution of the higher order terms is incorporated through the density gradients [25-27]. The second term in the expansion is the square gradient term that multiplies a parameter $\kappa_2$. The higher order gradient term captures the non-local behavior of the free energy. For the van der Waals model, the parameter $\kappa_2$ can be calculated using the following relation [31]:

$$\kappa_2 = -\frac{1}{6} \int_0^\pi r^2(-an(r))\, dr = \frac{a}{24\pi} \qquad (7)$$

This model is used later on to compare the density profiles while studying homogeneous nucleation with and without the square gradient term.

Using multibody DPD, the method has been primarily used to model 2D planar interfaces, drops, and two-phase oscillating cylinder simulations [24, 28-30]. The homogeneous nucleation of a vapor bubble in a bulk metastable phase has never been studied before using DPD. In this study, we use DPD to model the vapor bubble nucleation problem for the first time. First, the model parameters and the computational setup are described in Section II. This is followed by the results and discussion in section III and conclusion in section IV.

## II. MODEL PARAMETERS AND INITIAL SETUP

The isothermal DPD model described in [20-21] was used with the van der Waals multibody conservative force described in Eq. (6). The parameters chosen for the simulations are given in

Table 1. Systems were initialized at a metastable liquid state point $\overline{\rho}, \overline{k_B T}$ on the phase diagram, which was constructed for the van der Waals equation of state using the Maxwell equal area construction rule. The DPD beads were set up in a cubic domain with periodic boundary conditions in all three directions. The interaction radius of the particles is much larger than the size of the beads shown in this figure. All images were rendered using the visualization package Visual Molecular Dynamics (VMD) [32].

The systems were then allowed to evolve for 200,000 time steps with the data being stored at each time step. The van der Waals parameters in DPD units were chosen to be the same as the ones used by Tiwari et al. [30]. The 3D normalized Lucy function [33] was used as the weight function for the conservative forces and the simple weight function [20] was used for the dissipative and random forces. The integrator by Pagonabarraga et al. (SCPHF) described in [34] was used for all the simulations rather than the more widely used Groot-Warren integrator (GW) as it was found to be more stable in conserving temperature compared to the GW integrator for the cases of interest [35].

Before getting into the details of the results, it is important to discuss the relation between the local density $\overline{n}_i$ and the dimensionless density $\overline{\rho}$ that is used in DPD. The local density for a bead is defined as $\overline{n}_i = \sum_{j=1}^{N} \overline{w}(r_{ij})$ (assuming the normalization $[w] = 1$) and the dimensionless density as $\overline{\rho} = \frac{N}{\overline{\Omega}} \overline{r}_c^3$. The dimensionless density in DPD is the number of beads that can be found in a volume of radius $\overline{r}_c$ when there are $N$ beads in a volume of radius $\overline{L}$ i.e. $\overline{\rho} = \frac{N}{(\frac{4}{3}\pi\overline{L}^3)}(\frac{4}{3}\pi\overline{r}_c^3)$ and $\overline{\Omega} = \overline{L}^3$. The local density on the other hand is the sum of the normalized weight functions from all the neighbors that exist around each bead within the interaction range $\bar{r}_c$. The instantaneous local density, when time averaged over a sufficient time interval will equal the dimensionless density of the system. This was also confirmed in all the simulations that were done in this work.

## III. RESULTS AND DISCUSSION

Figure 1 shows the results of a simulation in which the DPD beads evolved from their initial single-phase metastable fluid state into two separate phases. The one distinguishing factor in these simulations is the presence of a characteristic 'void' in the middle of the domain shown as a blue 'sphere' in Fig. 1. The sphere in Fig. 1 was centered at the middle of the void. The void radius is set equal to the distance of the closest bead to the center of the void. However, this cannot possibly define a vapor bubble with a non-zero density. The void arises because of the visual representation of DPD beads. The beads in reality are coarse-grained particles with an interaction radius $\bar{r}_c$. However, while viewing them in VMD, the beads are shown as spherical particles with sizes that are not consistent with their interaction radius. In general, the DPD beads are coarse-grained particles with large overlapping volumes that fill up the entire domain and the reason the void is visible is due to the close-packing of the DPD interaction spheres as shown in Fig. 2.

**DEFINITION OF BUBBLE RADIUS AND IDENTIFICATION OF VAPOR REGION**

From inspection of Fig. 1, it is apparent that determination of local density variations and bubble radii from VMD images can be challenging. One way to better visualize a bubble and account for the density variation around it is to overlay a phantom grid in the domain and calculate the local density variation at each of these grid points. This phantom grid idea has been previously used in MD simulations to aid in the visualization of low density regions in bubble nucleation simulations [7]. A regular grid with spacing of 0.5 DPD units was used to calculate the local density at each of the grid points using the same weight function and interaction radius as the DPD force calculations. Once these points have been identified and tagged, the center of mass of this region can be found easily. This center of mass is also the center of the vapor (or bubble) region.

Once the center of the vapor region is found, the radial distribution of DPD beads around the void can be calculated. This helps in calculating the extent of the vapor region if the number of beads that make up the vapor region is known. One way to determine this is to use the continuum definition in which the bubble interface is defined as a region of zero mass and zero thickness with a step change in density. Although a gradual rather than sharp change in density across the interface is expected at the mesoscopic level, the continuum definition is useful for visualization purposes. This is indeed the case as we shall see in the next section. The volume of the vapor and liquid regions that are expected in a nucleation process can be calculated using the fact that the system is closed with fixed number of particles and a fixed system volume.

$$N = N_L + N_V \tag{8}$$

$$\bar{\Omega} = \bar{\Omega}_L + \bar{\Omega}_V \tag{9}$$

Here N is the total number of DPD beads in the system, $N_L$ is the number of beads in the liquid region, $N_V$ is the number of beads in the vapor region, $\bar{\Omega}$ is the total system volume, $\bar{\Omega}_L$ is the volume of the liquid region and $\bar{\Omega}_V$ is the volume of the vapor region. In DPD, the dimensionless density is defined as the number of particles contained in the volume $\bar{r}_c^3$.

Accordingly, the dimensionless densities $\bar{\rho}_L$ and $\bar{\rho}_V$ can be similarly defined for the saturated liquid and vapor regions as:

$$\bar{\rho}_L = \frac{N_L}{\bar{\Omega}_L} \bar{r}_c^3 \tag{10}$$

$$\bar{\rho}_V = \frac{N_V}{\bar{\Omega}_V} \bar{r}_c^3 \tag{11}$$

The dimensionless densities defined in Eqs. (10-11) are assumed to be the coexistence limits of the binodal curves as predicted by continuum theory and are known *a priori*. Using these equations, the volume of the vapor region and the number of beads that make up that vapor region can be calculated. Further assuming that the bubble is spherical the radius of the bubble can be calculated using this estimated volume. Table 2 shows the calculated values of volume and number of particles for the liquid and vapor regions and the resultant bubble radii for several cases. For the DPD simulations, the center of the bubble was calculated first using the phantom grid approach. After the center has been established, the time averaged distribution of beads was calculated in bins distributed radially from the center. Each radial bin was 0.05 DPD units thick as described above. In addition to continuum radius calculations, bubble radii were also computed using DPD particle distributions by assuming that the number of DPD beads in the bubble was equal to $N_V$. The exact location of the bin where the values matched was noted and its distance from the center of the bubble was calculated. This is listed as R-DPD in the last column of Table 2. The DPD simulation values show good agreement with continuum predictions. The closeness of the predicted and calculated radii gives confidence in the correctness of the model and the simulated results.

Once the radius of the bubble has been defined, the phantom grid can be used to understand the process of homogeneous bubble nucleation in a DPD system. The radius of the bubble can be used to define a threshold for the local density variation in the system. Figure 3 shows the local density variation from the center of the bubble for the system with 3375 beads and $\bar{r}_c = 4.0$. All grid points that fall below the threshold local density value are tagged as the vapor region and the rest as liquid.

Figure 4 shows the development of a low density vapor region in this system over time. The colored grid points are those that have been tagged as vapor. The snapshots are in increasing order of time starting initially with no grid points being tagged as vapor to the end where the vapor bubble just oscillates in space and time. The last frame is a top view of the bubble at the end of the simulation in the periodic domain. Initially, low density pockets appear all over the domain. However, they disappear over time. Smaller bubbles coalesce to form larger ones and the process continues until one region is formed that grows and remains stable for the rest of the simulation. This region oscillates over time, in shape and volume, with its center of mass moving throughout the domain. The vapor region appears to be disjointed in the frames, but that is the result of periodic boundaries. The visualization of the bubble using the phantom grid is definitely a convenient alternative method for capturing the vapor region explicitly once the bubble boundaries have been defined.

**INFLUENCE OF SYSTEM SIZE AND INTERACTION RADIUS ON LOCAL DENSITY VARIATION**

Figure 5 shows the dependence of local density on radial distance from the center of the bubble for different system sizes and at an interaction radius of 4.0. A bin size of 0.5 DPD units was used for these calculations. The local densities of all phantom grid points that lie within a particular bin were averaged to give the local density of the bin. The coexistence limits of the local density for the initial metastable state point at $\overline{k_BT} = 0.021$ and $\overline{\rho} = 27.0$ are also shown in the figure.

All the cases show a smooth variation of local density from the low-density vapor region to the high density liquid region. This is expected behavior as the sharp density profile is usually an approximation that exists at the continuum level [31]. One of the interesting features in both the figures is the low density limit on the vapor end for bigger system sizes. It is interesting to note that for the smallest system that was investigated $\overline{\Omega} = 10^3$, the limit is non-zero and matches well with the coexistence limits. But the deviation is more pronounced for the interaction radius of 4.0 and less for the interaction radius of 3.0 (not shown here).

The zero density limit at the end of the vapor region is an artifact of the void that exists in a DPD bubble and happens because of the close packing of the interaction spheres. To understand this close packing, let us look at Eq. (6). The conservative force is mainly a balance between two terms: the attractive term $\dfrac{\overline{b}\,\overline{k_BT}}{1-\overline{b}\overline{n}}$ and the repulsive term $\overline{a}$. At low densities where $\overline{b}\overline{n} \ll 1$, the attractive term $\overline{b}\,\overline{k_BT}$ is approximately equal to 5.25 x $10^{-4}$ DPD units and the repulsive term is 3.012 x $10^{-3}$ DPD units. The values from Table 1 were used to calculate these terms. At low densities, the repulsive terms are more dominant and cause the beads are repelled in the vapor phase. At high densities, the attractions are greater than the repulsions. The DPD beads are mesoscopic particles that can overlap and hence pack very closely in the liquid state due to strong attractions. The void is a consequence of the strong repulsions in the vapor regions and the size of the system.

The reason for the variation in slope evident in Fig. 5 for the different systems sizes is explained as follows. The size of the bubble scales with the size of the system. For smaller bubbles, the size of the void is small, but the interaction radius is a bigger fraction of the system size. For the same interaction radius, as the bubble size increases with system size, the size of the void also increases. Hence more of the void region is captured by the bins close to the center of the bubble, and the density profiles are sharper. The other interesting feature is the high density limit on the liquid end. The numerical values are higher than the coexistence limit at the liquid end. This indicates that the liquid beads close to the edges of the simulation box are more compressed than expected. However, this seems to only be a problem for larger domain sizes where a clearly visible void region is present as seen in Fig. 5 and Fig. 6.

In Fig. 6, the variations using different interaction radii are plotted for two system sizes. For the smaller system size of $\bar{\Omega} = 10^3$, close to the center of the bubble, the density is non-zero for all the cases even though it appears that for interaction radii of 2.0 and 3.0, the limit is close to, but greater than zero for all values of interaction radius. The magnitude of density is largest for the interaction radius of 4.0. In this case, the interaction is long range enough to capture interactions with the liquid and interface region. The numerical value is so small as to be obscured by the size of the points. If a much higher interaction radius of 4.0 is used, then the non-void region of the bubble is captured, which translates to the non-zero density for that case. On the liquid end, the interaction radius again has a noticeable, but inconsistent effect on the highest liquid density values attained. For locations farthest from the center of the bubble, the densities would be expected to be highest for the smallest interaction radii, because only the highest density regions are included in the interaction range. As $r_c$ increases, more and more of the vapor region would be included due to the small domain size and the periodic image of the system, thus lowering the density. This trend is observed for cases $r_c$=3 and $r_c$=4, but not for $r_c$=2. Using a bigger interaction radius would mean accounting for a larger number of beads from the periodic image of the system which would give much higher densities.

As the system size is increased in Fig. 6 to $\bar{\Omega} = 15^3$, the zero densities become prominent due to the bigger sizes of the void in the vapor region. The density variation does not show significant deviations with changes to the interaction radius in this case, although lower values of $r_c$ result in sharper interface profiles, as expected. The high densities on the liquid end also seem to approach the same values. The results in Fig. 5 and Fig. 6 indicate that the void and the liquid coexistence densities are sensitive to the ratio of interaction radius to the system size. Future studies, beyond the scope of the present work, will focus on performing comprehensive parametric studies to determine the most appropriate choice of interaction radii for given domain sizes. In any case, larger system sizes are preferable as they enable the establishment of a homogeneous liquid state for a given choice of interaction radius.

**INFLUENCE OF SQUARE-GRADIENT TERM ON LOCAL DENSITY VARIATION**

In DFT, the square-gradient term makes the free energy a non-local function of density. Figure 7 plots the density profile for all the system sizes for zero and non-zero value of $\kappa_2$. The addition of this term has no significant effect on the density variation from the bubble center. The variations are uniform and show the same features as the cases that were discussed before. One of the reasons why adding this term has no effect could be the use of a large interaction radius (with $\kappa_2 = 0$), which makes the effect of the conservative force and hence the free energy non-local. Hence addition of this non-local term does not contribute significantly to the interface region.

## IV. CONCLUSIONS

The liquid-vapor phase change model for liquid-vapor interfaces was studied in this work using the multibody framework based on the density functional theory of inhomogeneous fluids. The conservative force expression is used from previous work and is used to understand the problem of homogeneous nucleation in a periodic system. Tools for visualization of the bubble are developed using the phantom grid approach from MD studies. Using continuum thermodynamic ideas of interfaces, the extent of the vapor region is identified by calculating the number of beads that make up the vapor region. Once the vapor and liquid regions are clearly identified, visualization of the bubble is straightforward. The local density variation from the center of the bubble is studied for different system sizes and different interaction radii. The density variation was found to be smooth throughout. At the liquid end, higher than coexistence density values were found, which can be attributed to periodic boundaries and the higher compression of beads close to the edges. The square-gradient approximation to the free energy was also used but did not have any significant effect on the local density variation from the center of the bubble. The phase change model along with the visualization tools developed here can be used in tandem with the energy-conserving model from [36, 37] to study more complex phase-change problems in the future.

# ACKNOWLEDGMENT

This work was supported by the Office of Naval Research (grant no. N00014-07-1-0665).

**Fig. 1** Visualization of the 'void' and low density region in a 3375 bead simulation with $r_c = 4.0$. The 'void' region has been shown as a colored blue sphere in the middle of the domain

**Fig. 2** The figure shows the overlapping bead volumes present in a liquid-vapor system. a.) $r_c=4.0$, domain volume = $10^3$, b.) $r_c=4.0$, domain volume = $15^3$. The 'void' region in the center of the bubble for case b.) arises due to close-packing of the spheres around it. The vapor region extends beyond the void followed by the dense liquid region.

**Fig. 3** Local density variation for a 3375 bead system with interaction radius = 4.0 plotted radially from the center of the bubble. Figure shows the threshold used to distinguish vapor grid points from liquid ones based on the radius calculated from continuum theory

**Fig. 4** Visualization of the nucleation process in a system with 3375 beads and $r_c=4.0$ using the phantom grid; time is given in DPD units

**Fig. 5** Local density variation plotted radially from center of bubble for different system sizes and at an interaction radius of 4.0

**Fig. 6** Local density variation plotted radially from center of bubble for different interaction radii for two different system sizes

**Fig. 7** Local density variation plotted radially from center of bubble for interaction radius = 4.0 with kappa2 = 0 and kappa2 = a/(24*pi)

**Table 1 Values of dimensionless DPD parameters (in DPD units) that were held constant for all the liquid-vapor simulations**

**Table 2 Continuum predictions of radius of vapor bubble and comparisons with DPD simulations; the coexistence densities are $\overline{\rho}_L = 31.14$, $\overline{\rho}_V = 2.08$ for the initial metastable state point at $\overline{k_BT} = 0.021$ and $\overline{\rho} = 27.0$**

**\* These values have been rounded to the nearest integer**

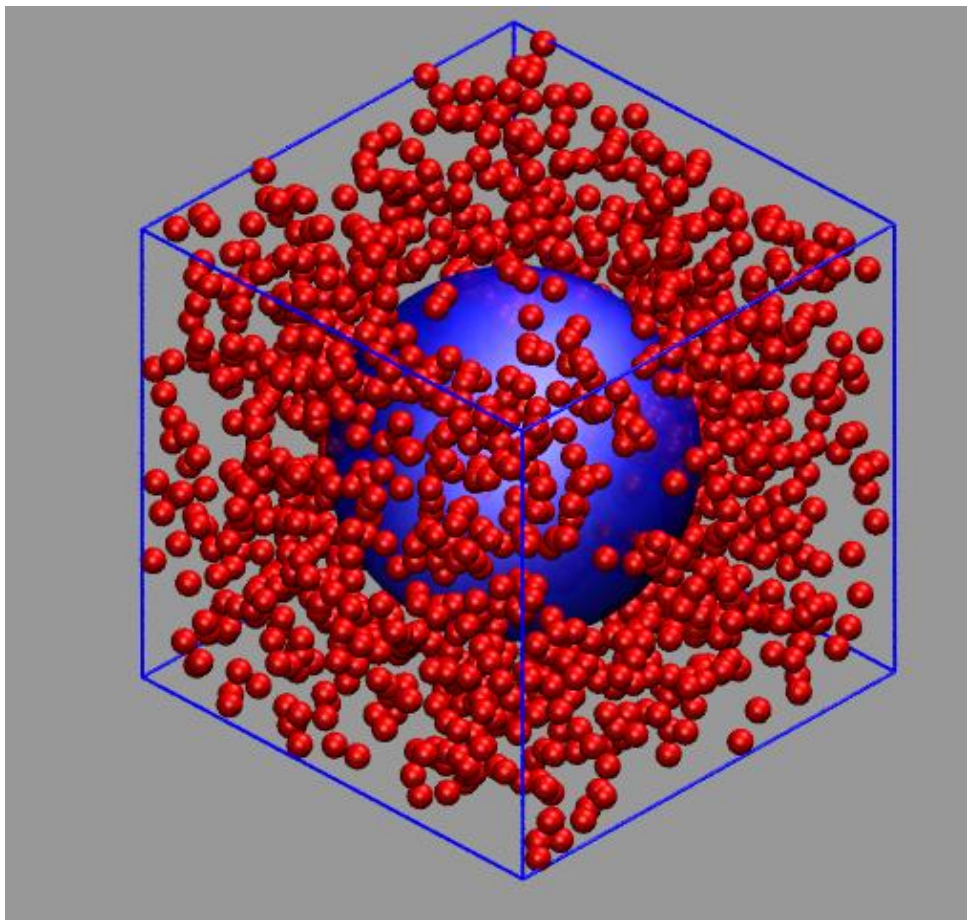

**Fig. 1.**

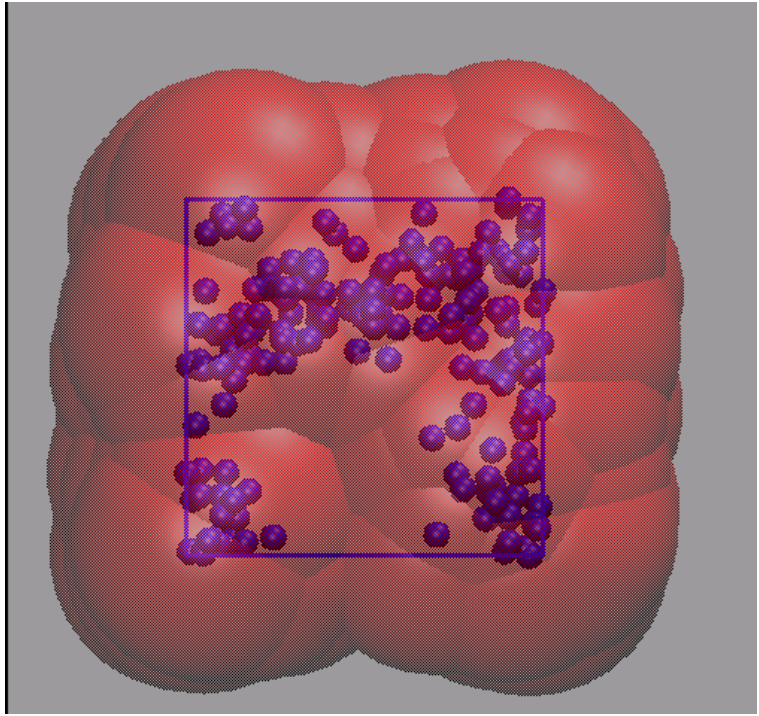

**a.)**

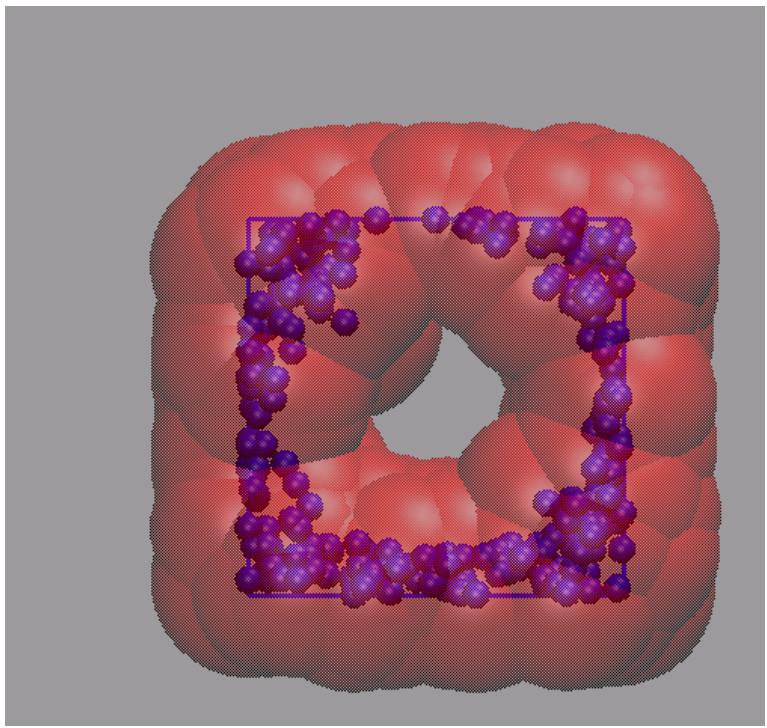

**Fig. 2. b.)**

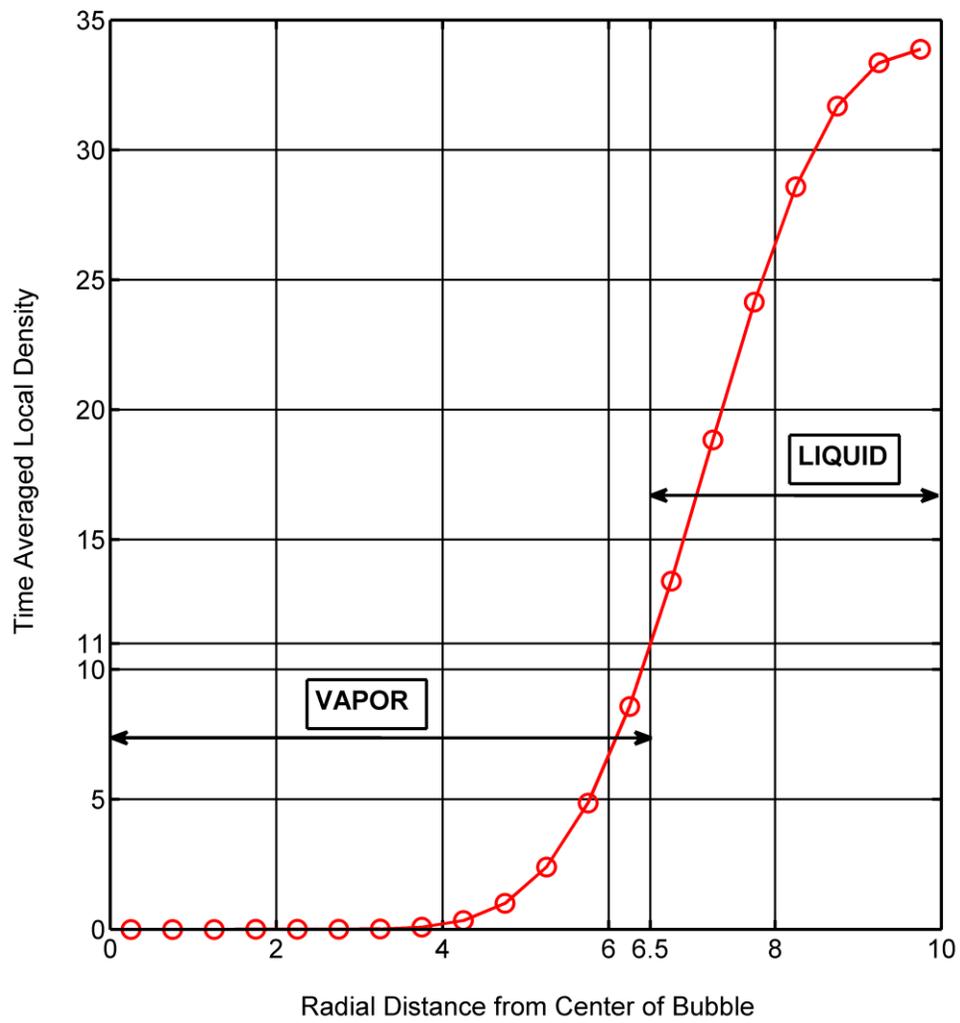

**Fig. 3.**

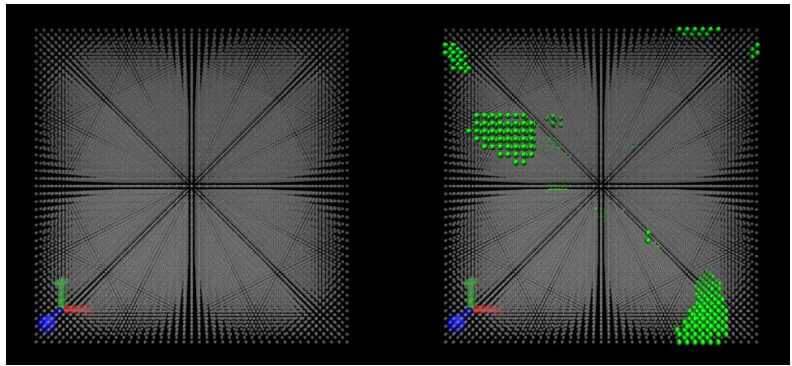

(a) t =0     (b) t = 20

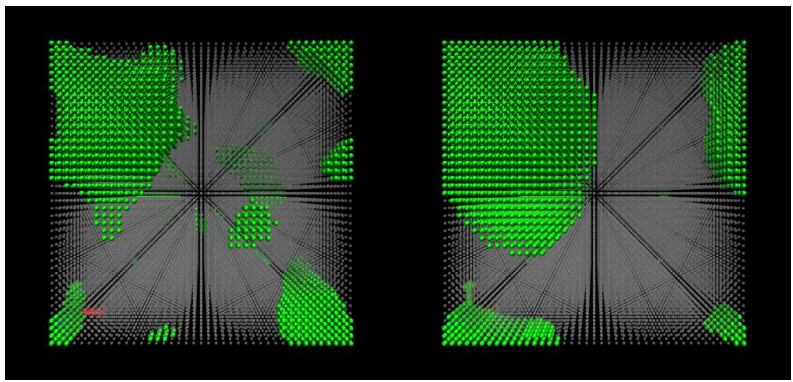

(c) t =100     (d) t = 1000

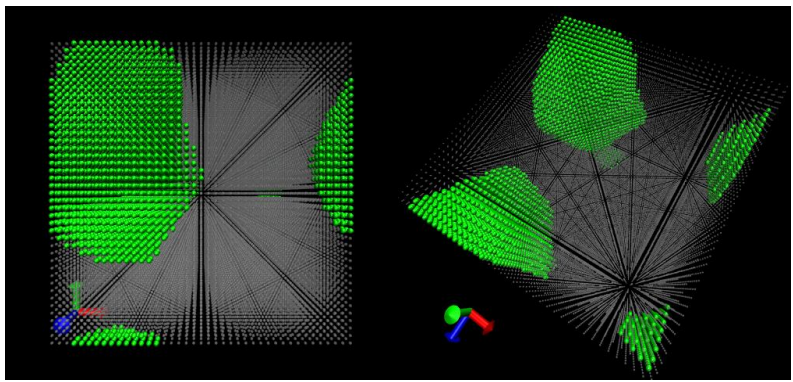

(c) t =1980     (d) t = 1980, rotated view

**Fig. 4.**

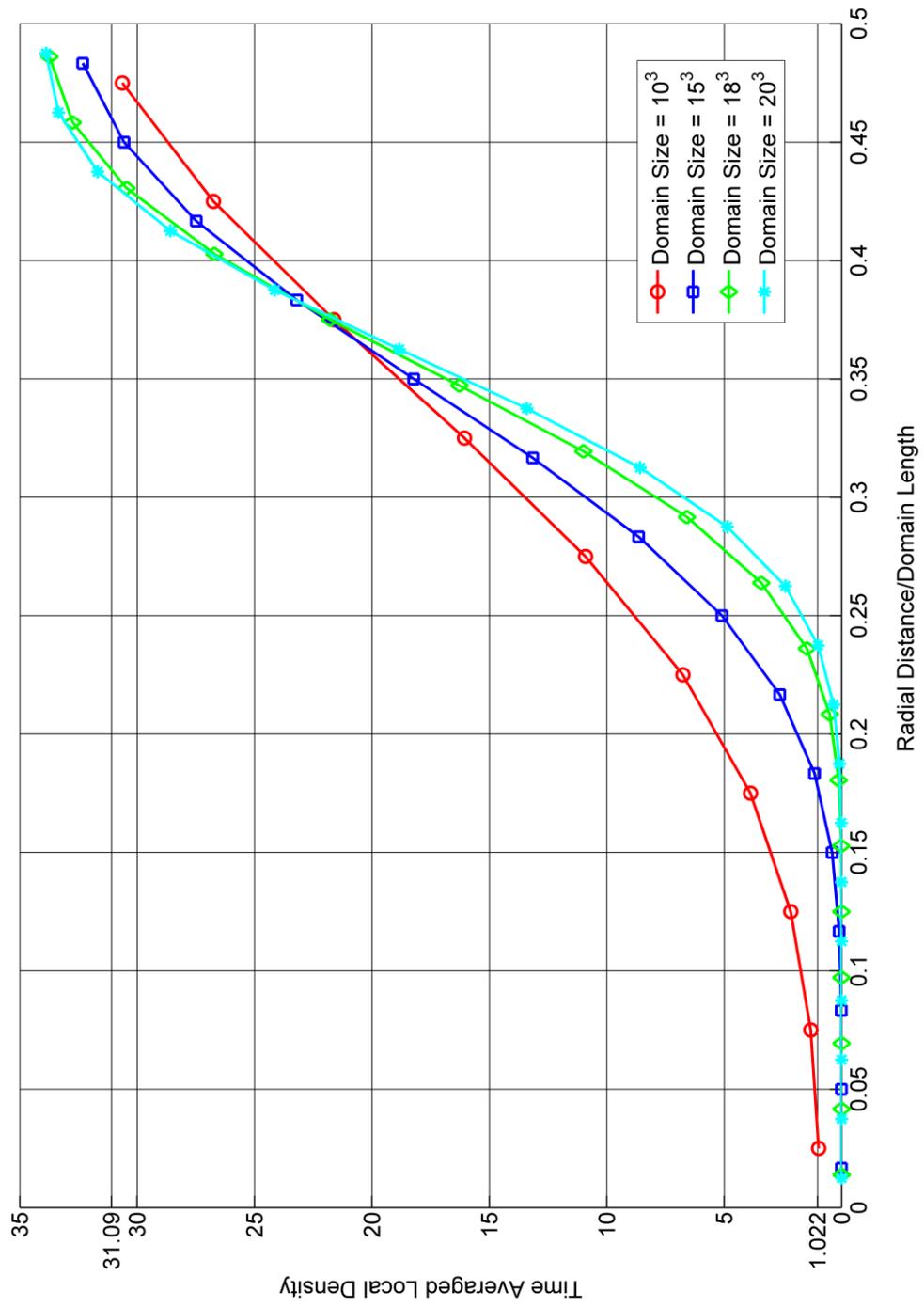

**Fig. 5.**

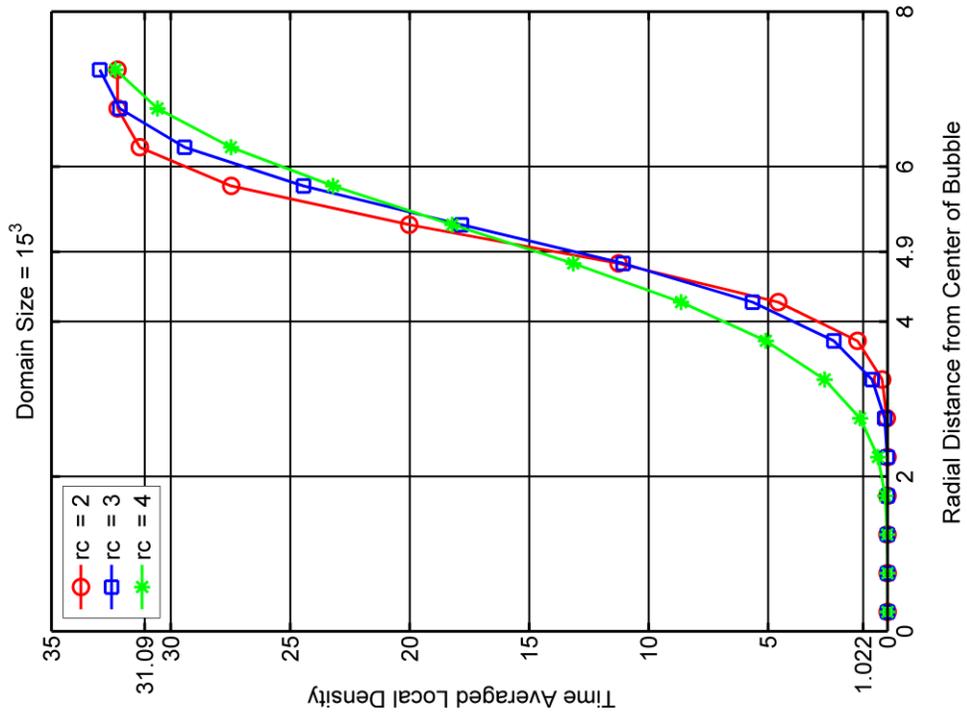
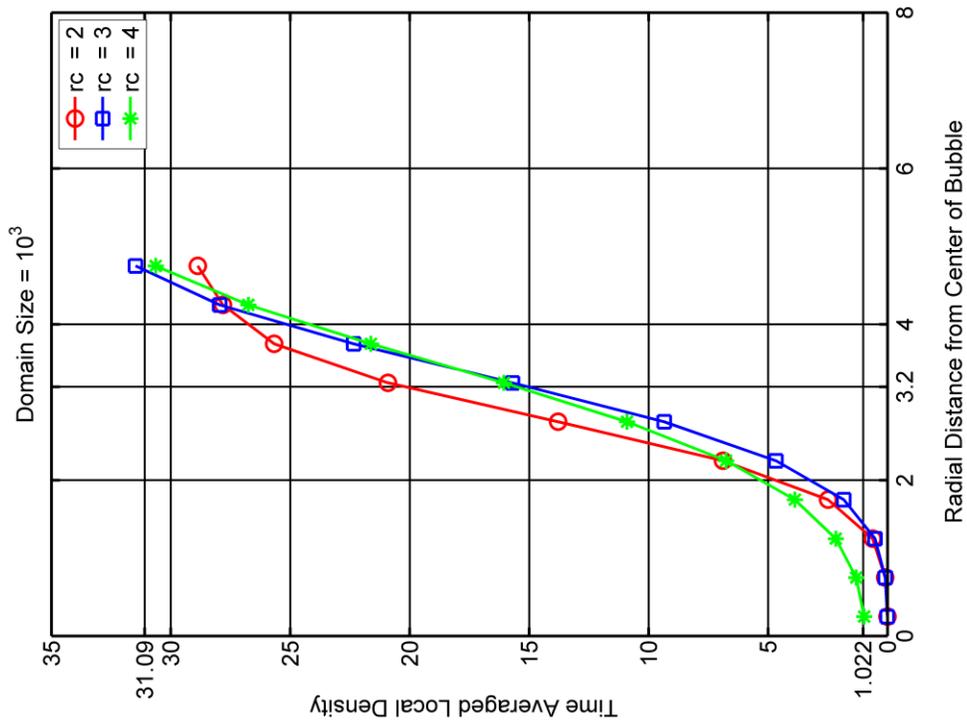

**Fig. 6.**

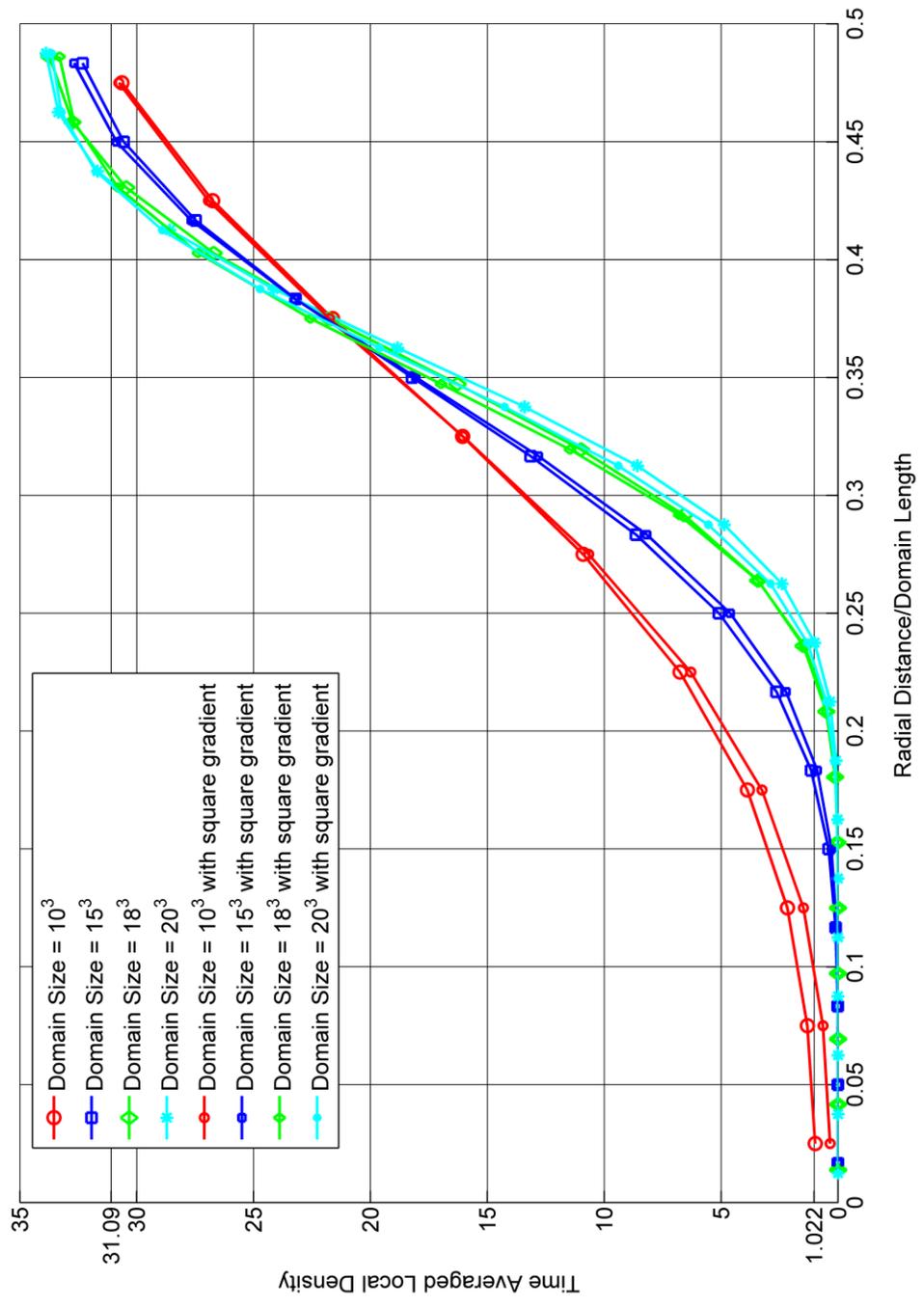

**Fig. 7.**

**Table 1.**

| PARAMETER | DESCRIPTION | VALUE |
|---|---|---|
| $\overline{k_B T}$ | Temperature | 0.021 |
| $\overline{\rho}$ | Dimensionless density | 27.0 |
| $\overline{\sigma}$ | Noise (random) parameter | 0.205 |
| $\overline{\gamma}$ | Friction (dissipative) parameter | $\overline{\gamma} = \dfrac{\overline{\sigma}^2}{2\overline{k_B T}}$ |
| $\Delta \overline{t}$ | Time step | 0.01 |
| $\overline{a}$ | van der Waals constant | 0.003012 |
| $\overline{b}$ | van der Waals constant | 0.025 |
| $N$ | Number of DPD beads | See Table 2 |
| $\overline{\Omega}$ | System volume | See Table 2 |
| $N_{\text{steps}}$ | Number of iterations | See Table 2 |

**Table 2.**

| SET PARAMETERS | | | CALCULATED CONTINUUM | | | | | |
|---|---|---|---|---|---|---|---|---|
| $\bar{r}_c$ | $N$ | $\bar{\Omega}$ | $\bar{\Omega}_V$ | $\bar{\Omega}_L$ | $N_V$* | $N_L$* | R Continuum | R DPD |
| 3.0 | 1000 | $10^3$ | 142.5 | 857.5 | 11 | 989 | 3.2 | 3.3 |
|  | 3375 | $15^3$ | 480.8 | 2894.2 | 37 | 3338 | 4.9 | 5.2 |
|  | 5832 | $18^3$ | 830.5 | 5001.5 | 64 | 5768 | 5.8 | 6.1 |
|  | 8000 | $20^3$ | 1139.7 | 6860.3 | 88 | 7912 | 6.5 | 6.9 |
| 4.0 | 422 | $10^3$ | 142.5 | 857.5 | 5 | 417 | 3.2 | 3.6 |
|  | 1424 | $15^3$ | 480.8 | 2894.2 | 16 | 1408 | 4.9 | 5.0 |
|  | 2460 | $18^3$ | 830.5 | 5001.5 | 27 | 2433 | 5.8 | 6.4 |
|  | 3375 | $20^3$ | 1139.7 | 6860.3 | 37 | 3338 | 6.5 | 7.1 |